# Neural dynamics under active inference: plausibility and efficiency of information processing

Lancelot Da Costa[1,2], Thomas Parr[2], Biswa Sengupta[2,3,4], Karl Friston[2]

[1]*Department of Mathematics, Imperial College London*
[2]*Wellcome Centre for Human Neuroimaging, Institute of Neurology, University College London*
[3]*Core Machine Learning group, Zebra AI, London*
[4]*Department of Bioengineering, Imperial College London*

Correspondence: Lancelot Da Costa l.da-costa@imperial.ac.uk

**Abstract**

Active inference is a normative framework for explaining behaviour under the free energy principle – a theory of self-organisation originating in neuroscience. It specifies neuronal dynamics for state-estimation in terms of a descent on (variational) free energy – a measure of the fit between an internal (generative) model and sensory observations. The free energy gradient is a prediction error – plausibly encoded in the average membrane potentials of neuronal populations. Conversely, the expected probability of a state can be expressed in terms of neuronal firing rates. We show that this is consistent with current models of neuronal dynamics and establish face validity by synthesising plausible electrophysiological responses. We then show that these neuronal dynamics approximate *natural gradient descent*, a well-known optimisation algorithm from information geometry that follows the steepest descent of the objective in information space. We compare the information length of belief updating in both schemes, a measure of the distance travelled in information space that has a direct interpretation in terms of metabolic cost. We show that neural dynamics under active inference are metabolically efficient and suggest that neural representations in biological agents may evolve by approximating steepest descent in information space towards the point of optimal inference.

## Introduction

Active inference is a normative framework for explaining behaviour under the free energy principle, a theory of self-organisation originating in neuroscience [1–4] that characterises certain systems at steady-state as having the appearance of sentience [5,6]. Active inference describes agents' behaviour as following the equations of motion of the free energy principle so as to remain at steady-state, interpreted as the agent's goal [7].

Active inference describes organisms as inference engines. This assumes that organisms embody a generative model of their environment. The model encodes how the states external to the agent influence the agent's sensations. Organisms infer their surrounding environment from sensory data by inverting the generative model through minimisation of variational free energy. This corresponds to performing approximate Bayesian inference (also known as variational Bayes) [2,3,8–11] or minimising the discrepancy between predictions and sensations [1,12]. Active inference unifies many existing theories of brain function [13,14], such as, for example, optimal control [15–18], the Bayesian brain hypothesis [19–21] and predictive coding [19,22–24]. It has been used to simulate a wide range of behaviours in neuropsychology, machine learning and robotics. These include planning and navigation [25–31], exploration [32–36], learning of self and others [37,38], concept learning [39–41], playing games [42–44], adapting to changing environments [45–47], active vision [48–51] and psychiatric illness [42,52–55]. Active inference agents show competitive or state-of-the-art performance in a wide variety of simulated environments [34,46,47,56].

The last decade has developed a theory of how the brain might be implementing active inference consistently with neurology [1,7,49,57,58]. A small number of predictions of this theory have been empirically validated, including the role of dopamine in decision-making [59,60] and free energy minimisation in exploration and choice behaviour [61,62]. This paper investigates the neural dynamics of this process theory from two complementary standpoints. 1) Consistency with empirically driven models of neural population dynamics. 2) And the metabolic and computational efficiency of such dynamics.

Efficiency is an important aspect of neural processing in biological organisms [63–68] and an obvious desideratum for artificial agents. Efficiency of neural processing in biological agents is best seen in the efficient coding hypothesis by Horace Barlow [69–71], which has received much empirical support [65,66,72–77] (see in particular [64] for energetic efficiency) and has been a successful driver in computational neural modelling [75,78–80]. In brief, any process theory of brain function should exhibit reasonably efficient neural processing.

Active inference formalises perception as inferring the state of the world given sensory data through minimisation of variational free energy. This amounts to constantly optimising Bayesian beliefs about states of the outside world in relation to sensations. By beliefs, we mean probability distributions over states of the environment. These distributions score the extent to which an agent trusts that the environment is, or not, in this or another state. From an information theoretic viewpoint, a change of beliefs is a computation or a change in information encoded by the agent.

This belief updating has an associated energetic cost. Landauer famously observed that a change in information entails heat generation [81,82]. It follows that the energetics needed for a change in beliefs may be quantified by the change in information encoded by the agent over time (as the organism has to alter, e.g., synaptic weights or restore transmembrane potentials) mathematically scored by the length of the path travelled by the agent's beliefs in information space. There is a direct correspondence between the (Fisher) information length of a path and the energy consumed by travelling along that path [83,84]. To ensure metabolic and computational efficiency, an efficient belief updating algorithm should reach the free energy minimum (i.e., the point of optimal inference) via the shortest possible path on average. Furthermore, since an agent does not know the free energy minimum in advance, she must find it using only local information about the free energy landscape. This is a non-trivial problem. Understanding how biological agents solve it might not only improve our understanding of the brain, but also yield useful insights into mathematical optimisation and machine learning.

In the first part of this work, we show that the dynamics prescribed by active inference for state-estimation are consistent with current models of neural population dynamics. We then show that these dynamics approximate natural gradient descent on free energy, a well-known optimisation algorithm from information geometry that follows the steepest descent of the objective in information space [85]. This leads to short paths for belief updates as the free energy encountered in discrete state-estimation is convex (see Appendix A). These results show that active inference prescribes efficient and biologically plausible neural dynamics for state-estimation and suggest that neural representations may be collectively following the steepest descent in information space towards the point of optimal inference.

## The softmax activation function in neural population dynamics

This section rehearses a basic yet fundamental feature of mean-field formulations of neural dynamics; namely, the average firing rate of a neural population follows a sigmoid function of the average membrane potential. It follows that firing rates can be expressed as a softmax function of average transmembrane potentials, when considering multiple coupled neural populations, as the softmax is simply a generalisation of the sigmoid to vector inputs. This can be seen by the fact that the sigmoid— respectively softmax— function is used in univariate— respectively multivariate— logistic regression.

The sigmoid relationship between membrane potential and firing rate, was originally derived by Wilson and Cowan [86], who showed that any unimodal distribution of thresholds within a neural population, whose individual neurons are modelled as a Heaviside response unit, results in a sigmoid activation function at the population level. This is because the population’s activation function equals a smoothing (i.e., a convolution) of the Heaviside function with the distribution of thresholds.

The assumption that the sigmoid arises from the distribution of thresholds in a neural population remained unchallenged for many years. However, the dispersion of neuronal thresholds is, quantitatively, much less important than the variance of neuronal membrane potential within populations [87]. Marreiros and colleagues showed that the sigmoid activation function can be more plausibly motivated by considering the variance of neuronal potentials within a population [88], which is generally modelled by a Gaussian distribution under the Laplace assumption in mean-field treatments of neural population dynamics [89]. Briefly, with a low variance on neuronal states, the sigmoid function that is obtained – as a convolution of the Heaviside function – has a steep slope, which means that the neural population as a whole, fires selectively with respect to the mean membrane potential, and vice-versa. This important fact, which was verified experimentally using dynamic causal modelling [88,90], means that the variance of membrane potentials implicitly encodes the (inverse) precision of the information encoded within the population.

Currently, the sigmoid activation function is the most commonly used function to relate average transmembrane potential to average firing rate in mean-field formulations of neural population dynamics [91,92] and deep neural networks [93,94]. This relationship logically extends to a softmax function when considering multiple coupled neural populations.

## Neural dynamics of perceptual inference

Active inference formalises perception as inferring the state of the world given sensory data through minimisation of variational free energy [7]. For state estimation on discrete state-space generative models (e.g., partially observable Markov decision processes [95]), the free energy gradient corresponds to a generalised a prediction error [7]. This means that to infer the states of their environment, biological agents reduce the discrepancy between their predictions of the environment and their observations, or maximise the mutual information between them [63].

Variational free energy is a function of approximate posterior beliefs $Q$,

$$\begin{aligned}
\mathrm{F}(Q) &\triangleq E_{Q(s)}\left[\ln Q(s) - \ln P(o,s)\right] \\
&= \underbrace{D_{KL}\left[Q(s) \,\|\, P(s \mid o)\right]}_{\geq 0} - \underbrace{\ln P(o)}_{\text{Log-evidence}} \\
&= \underbrace{D_{KL}\left[Q(s) \,\|\, P(s)\right]}_{\text{Complexity}} - \underbrace{E_{Q(s)}\left[\ln P(o \mid s)\right]}_{\text{Accuracy}}
\end{aligned}$$

while $P$ is the generative model: a probability distribution over hidden states ($s$) and observations ($o$) that encodes the causal relationships between them. Only the observations are directly accessible; hidden states can only be inferred. The symbol $E_Q$ means the expectation (i.e., the average) of its argument under the subscripted distribution. $D_{KL}$ is known as the Kullback-Leibler divergence or relative entropy [96–98] and is used as a non-negative measure of the discrepancy between two probability distributions. Note that this is not a measure of distance, as it is asymmetric. The second line here shows that minimising free energy amounts to approximate the posterior over hidden states, which is generally intractable to compute directly, with the approximate posterior. Exact Bayesian inference requires the approximate and true posterior to be exactly the same, at which point free energy becomes negative log model evidence (a.k.a., marginal likelihood). This explains why the (negative) free energy is sometimes referred to as an evidence lower bound (ELBO) in machine learning [93]. The final line shows a decomposition of the free energy into accuracy and complexity, underlying the need to find the most accurate explanation for sensory observations that is minimally complex (c.f., Horace Barlow's principle of minimum redundancy [69]).

When a biological organism represents some of its environment in terms of a finite number of possible states (e.g., the locations in space encoded by place cells), we can specify its belief updating about the current state in peristimulus time. Discrete state-estimation is given as a (softmax) function of accumulated negative free energy gradients [57].

$$\dot{v} = -\nabla_{\mathbf{s}} F$$
$$\mathbf{s} = \sigma(v)$$

In this equation, $\sigma$ is a softmax function and $\mathbf{s}$ represents the agent's beliefs about states. (These are the parameters of a categorical distribution $Q$ over states). Explicitly, $\mathbf{s}$ is a vector whose *i*-th component is the agent's confidence (expressed as a probability) that it is in the *i*-th state. The softmax function is the natural choice to map from free energy gradients to beliefs as the former turns out to be a logarithm [7] and the components of the latter must sum to one.

Just as neuronal dynamics involve translation from post-synaptic potentials to firing rates, these dynamics involve translating from a vector of real numbers ($v$), to a vector where components are bounded between zero and one ($\mathbf{s}$). As such, we can interpret $v$ as the voltage potential of neuronal populations, and $\mathbf{s}$ as representing their firing rates (since these are upper bounded thanks to neuronal refractory periods). Note the softmax function here plays the same role as in mean-field formulations; it translates average potentials to firing rates. On the one hand, this view is consistent with models of neuronal population dynamics. On the other hand, it confers post-hoc face validity, as it enables to synthesise plausible local field potentials (see Figure 1) and a wide range of other electrophysiological responses, including repetition suppression, mismatch negativity, violation responses, place-cell activity, phase precession, theta-gamma coupling, and more [57].

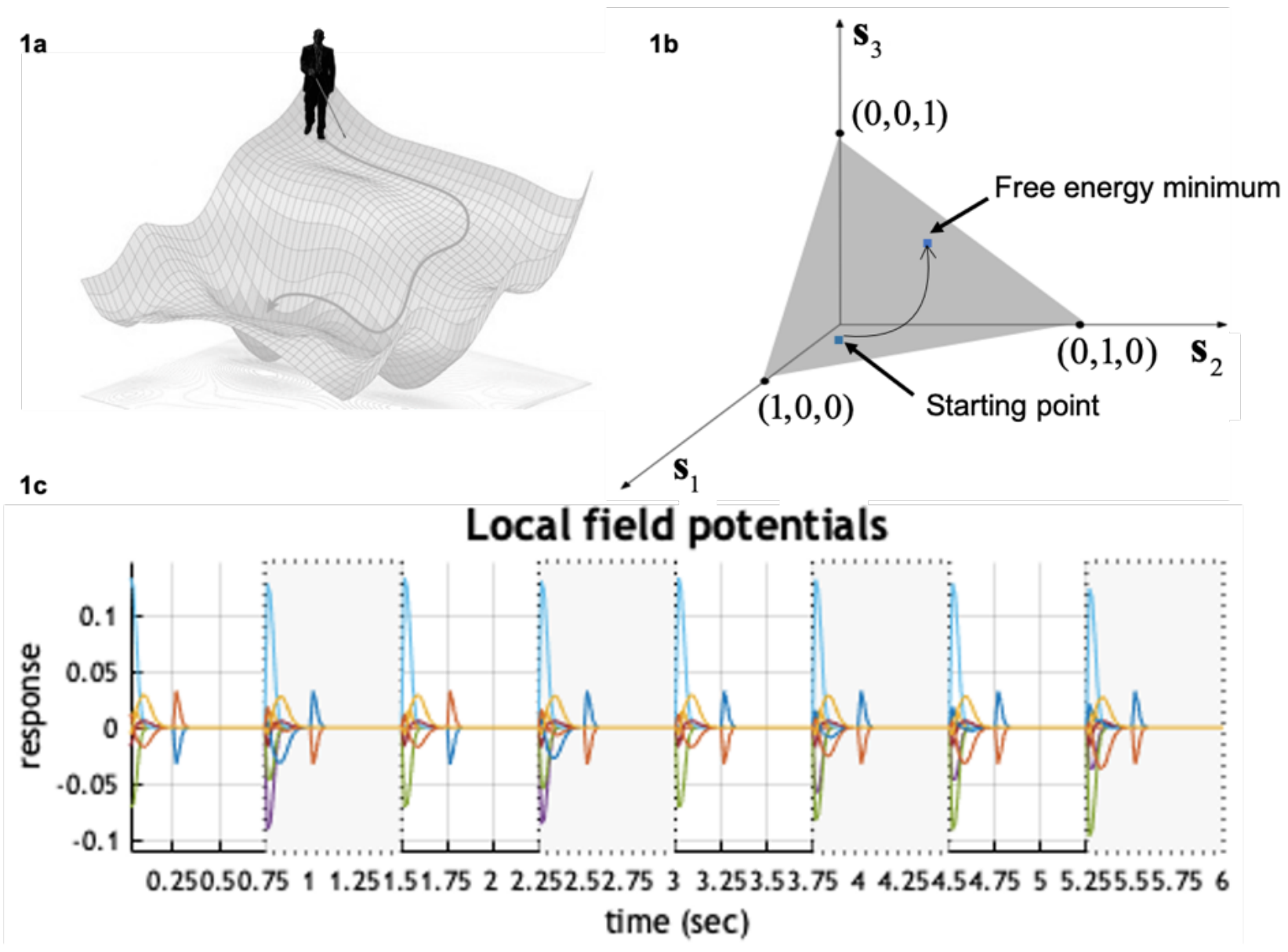

**Figure 1: Active inference for state estimation (discrete state-space).**

*Panel 1a* summarises the problem of finding the minimum of a function (e.g., the free energy). One possibility would be taking the shortest path, which involves climbing up a hill, however in the nescience of the minimum, a viable strategy consists of myopically taking the direction of steepest descent. In *panel 1b*, we depict an example of a trajectory of an agent's beliefs during the process of perception, which consists of updating beliefs about the states of the external world to reach the point of optimal inference (i.e., free energy minimum). In this example the state-space comprises only three states (e.g., three different locations in a room). As they are probabilities over states, the components of $\mathbf{s}$ are non-negative and sum to one; hence, the agent's beliefs naturally live on a triangle in three-dimensional space. Mathematically, this object is called a (two-dimensional) *simplex.* This constitutes the belief space, or information space, on which the free energy is defined. Technically, this object is a smooth statistical manifold, which corresponds to the set of parameters of a categorical distribution. To optimise metabolic and computational efficiency, agents must update their beliefs to reach the free energy minimum via the shortest possible path on this manifold. In *panel 1c* we exhibit simulated local field potentials that arise by interpreting the rate of change of $v$ in terms of depolarisations, over a sequence of eight observations (e.g., saccadic eye-movements). As the rate of change is given by the free energy gradients, the decay of these local field potentials to zero coincides with reaching the free energy minimum (at which the gradient is zero by definition). These were obtained during the first numerical simulation described in Figure 3. For more details on the generation of simulated electrophysiological responses, see [57].

The idea that state-estimation is expressed in terms of firing rates is well-established when the state-space constitutes an internal representation of space. This is the raison d'être of the study of place cells [99], grid cells [100] and head-direction cells [101,102], where the states inferred are physical locations in space. Primary afferent neurons in cats have also been shown to encode kinematic states of the hind limb [103–105]. Most notably, the seminal work of Hubel and Wiesel [106] showed the existence of neurons encoding orientation of visual stimuli. In short, the very existence of receptive fields in neuroscience speaks to a carving of the world into discrete states under an implicit discrete state generative model. While many of these studies focus on single neuron recordings, the arguments presented above are equally valid and generalise the case of 'populations' comprising of a single neuron.

In summary, the neuronal dynamics associated with state-estimation in active inference are consistent with mean-field models of neural population dynamics. This view is strengthened a posteriori, as this allows one to generate a wide range of plausible electrophysiological responses. Yet, further validation remains to be carried out by testing these electrophysiological responses empirically. We will return to this in the discussion.

## A primer on information geometry and natural gradient descent

To assess the computational and metabolic efficiency of a belief trajectory, it becomes necessary to formalise the idea of 'belief space'. These are well-studied structures in the field of information geometry [107–109], called statistical manifolds. In our case, these are (smooth) manifolds, where each point corresponds to a parameterisation of the probability distribution in consideration (see Figure 2). One is then licensed to talk about a change in beliefs as a trajectory on a statistical manifold.

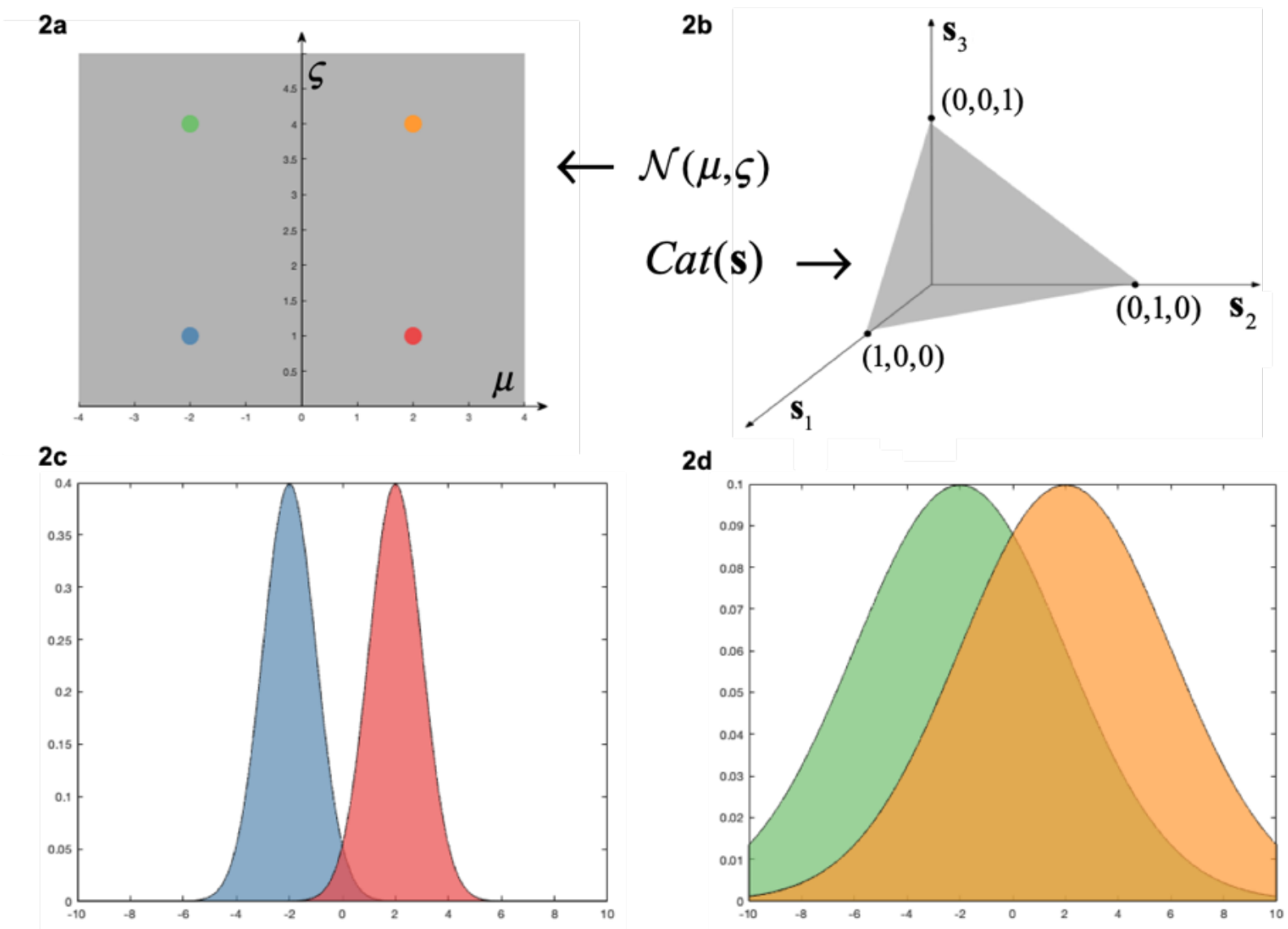

**Figure 2: Statistical manifolds and information length.**

*Panels 2a-b* illustrate the statistical manifolds associated with two well-known probability distributions: the normal distribution and the categorical distribution, respectively. The statistical manifold associated with a probability distribution is the set of all possible parameters that it can take. For the univariate normal distribution, parameterised with mean $\mu$ and positive standard deviation $\varsigma$, the associated statistical manifold is the upper half plane (*panel 2a*). For the categorical distribution, in the case of three possible states, the statistical manifold is the 2-dimensional simplex (*panel 2b*). More generally, in the case of *n* possible states, the statistical manifold of the categorical distribution is the set of all vectors with positive components that sum to one, i.e., the (n-1)-dimensional simplex. This is a higher-dimensional version of the triangle or the tetrahedron. In *panels 2c-d* we illustrate why the Euclidean distance is ill-suited to measure the information distance between probability distributions. To show this we selected four distributions that correspond to points on the statistical manifold of the normal distribution. One can see that the Euclidean distance between the modes of the red and the blue distributions is the same as that from the orange and the green, however, the difference in information of each respective pair is quite different. In *panel 2c*, the two distributions correspond to two drastically different beliefs, since there is such little overlap; on the contrary, the beliefs in *panel 2d* are much more similar. This calls for a different notion of distance that measures the difference in (Fisher) information between distributions; namely, the information length.

Smooth statistical manifolds are naturally equipped with a different notion of distance, even though they may be subsets of Euclidean space. This is because the Euclidean distance measures the physical distance between points, while the information length measures distance in terms of the (accumulated) change in (Fisher) information (see Figure 2) along a path. The canonical choice of information length on a statistical manifold is associated with the Fisher information metric tensor $\mathbf{g}$ [110–112]. Technically, a metric tensor is a smoothly varying choice of symmetric, positive definite matrix at each point of the manifold. This enables computation of the length of paths as well as the distance between points, by measuring the length of the shortest path (see Appendix B). Mathematically, the Fisher information metric can be defined the Hessian of the KL divergence between two infinitesimally close distributions (see Appendix B). This means that the information length of a trajectory on a statistical manifold is given by accruing infinitesimally small changes in the KL divergence along it.

Amari's natural gradient descent [85,113] is a well-known optimisation algorithm for finding the minimum of functions defined on statistical manifolds such as the variational free energy. It consists of preconditioning the vanilla gradient descent update rule with the inverse of the Fisher information metric tensor:

$$\dot{\mathbf{s}} = -\nabla_{\mathbf{s}} \mathrm{F} \quad \rightarrow \quad \dot{\mathbf{s}} = -\mathbf{g}^{-1}(\mathbf{s})\nabla_{\mathbf{s}} \mathrm{F}$$

Preconditioning by the inverse of $\mathbf{g}$ means that the natural gradient privileges directions of low information length. One can see this, since the directions of greatest (resp. smallest) information length are the eigenvectors of the highest (resp. lowest) eigenvalues of $\mathbf{g}$.

In fact, Amari proved that the natural gradient follows the direction of steepest descent of the objective function in information space [85]. Technically, the natural gradient generalises gradient descent to functions defined on statistical manifolds. As the free energy for discrete state-estimation is convex (see Appendix A), this means that natural gradient descent will always converge to the free energy minimum via a short path.

To summarise, agents' beliefs naturally evolve on a statistical manifold towards the point of optimal inference. These manifolds are equipped with a different notion of distance; namely, the information length. In the case of discrete state-estimation, beliefs evolve on the simplex towards the free energy minimum. Reaching the minimum with a short path translates into higher computational and metabolic efficiency. One scheme that achieves short paths for finding the minimum of the free energy is the natural gradient. In the next section, we will show that the neuronal dynamics entailed by active inference approximate natural gradient descent.

## Active inference approximates natural gradient descent

Discretising the (neuronal) dynamics prescribed by active inference and natural gradient descent give us the following state-estimation belief updates, respectively:

$$\mathbf{s}^{(t+1)} \leftarrow \sigma\left(\ln \mathbf{s}^{(t)} - \epsilon \, \nabla_{\mathbf{s}^{(t)}} \mathrm{F}\right) \qquad \mathbf{s}^{(t+1)} \leftarrow \frac{\mathbf{s}^{(t)} - \epsilon \, \mathbf{g}^{-1}(\mathbf{s}^{(t)})\nabla_{\mathbf{s}^{(t)}} \mathrm{F}}{\sim}$$

In these equations the logarithm is taken component-wise, $\varepsilon$ is the step-size used in the discretisation and $\sim$ denotes normalisation by the sum of the components, to ensure that $\mathbf{s}^{(t+1)}$ lies on the simplex[1].

[1] Natural gradient descent dynamics do not necessarily remain on the statistical manifold. This problem has been the object of numerous works that supplemented the natural gradient update with a projection step [114–118]. Here we choose the simplest projection step to ensure that the result remains on the simplex: normalising with the sum of the components.

These dynamics are approximately the same. We can equate them under a first order Taylor approximation of the exponential inside the softmax function:

$$
\begin{aligned}
\mathbf{s}^{(t+1)} &\leftarrow \sigma\left(\ln \mathbf{s}^{(t)} - \epsilon\, \nabla_{\mathbf{s}^{(t)}} \mathrm{F}\right) \\
&= \frac{\exp\left[\ln \mathbf{s}^{(t)} - \epsilon\, \nabla_{\mathbf{s}^{(t)}} \mathrm{F}\right]}{\sim} \\
&= \frac{\mathbf{s}^{(t)} \odot \exp\left[-\epsilon\, \nabla_{\mathbf{s}^{(t)}} \mathrm{F}\right]}{\sim} \\
&\simeq \frac{\mathbf{s}^{(t)} \odot \left[\vec{1} - \epsilon\, \nabla_{\mathbf{s}^{(t)}} \mathrm{F}\right]}{\sim} \\
&= \frac{\mathbf{s}^{(t)} - \epsilon\, \mathbf{s}^{(t)} \odot \nabla_{\mathbf{s}^{(t)}} \mathrm{F}}{\sim} \\
&= \frac{\mathbf{s}^{(t)} - \epsilon\, \mathbf{g}^{-1}(\mathbf{s}^{(t)}) \nabla_{\mathbf{s}^{(t)}} \mathrm{F}}{\sim}
\end{aligned}
$$

The symbol $\odot$ denotes the Hadamard product (elementwise multiplication). The last line follows since, on the simplex, the inverse of the Fisher information metric tensor is simply a diagonal matrix whose diagonal is $\mathbf{s}$ (see Appendix C).

Although these dynamics are approximately the same, this does not guarantee that the paths taken in the limit of infinitessimaly small time steps (which correspond to continuous-time dynamics in physical and biological systems) will be the same. One can see this algebraically, since the number of time steps needed to reach the free energy minimum increases as the step-size decreases, thus the difference between paths, which equals the sum of the differences at each timestep, is not guaranteed to converge to zero. Hence it is necessary to verify that this approximation holds well in practice, by analysing the discrepancy between paths using numerical simulations.

## Numerical simulations

In this section we use numerical simulations of two tasks (i.e., a two-step maze and a rule learning task) to assess to what extent the correspondence between active inference and natural gradient descent holds true. Our simulations compared the information length of the belief trajectories taken by both schemes which reflects their computational and metabolic efficiency. See Figure 3 for details.

We observe that both schemes perform equally well on average across both tasks. It is interesting to see that in some cases, the trajectories taken by both schemes are significantly longer than the shortest path (i.e., the geodesic; see Appendix D). This is unsurprising since agents' beliefs evolve towards the free energy minimum using only local information about the free energy landscape. Note that this occurred only in a minority of the trials in the examples considered here.

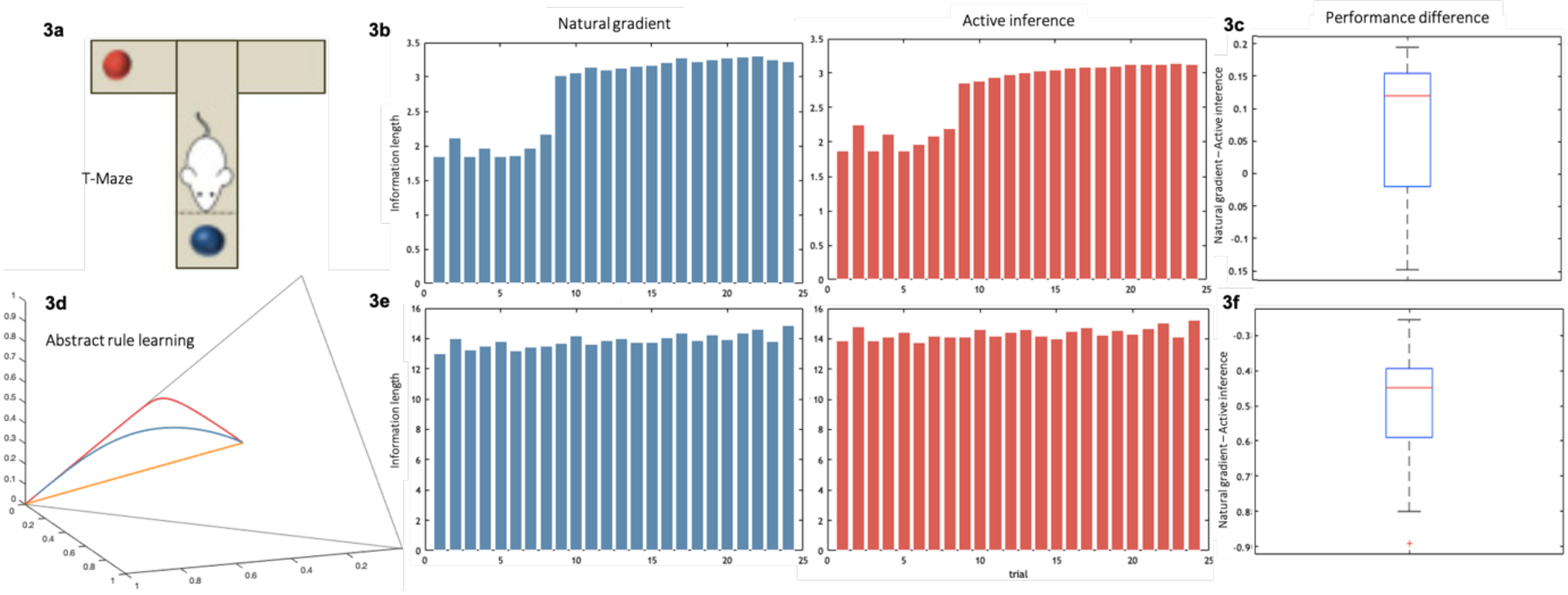

**Figure 3: Information length and belief trajectories of active inference and natural gradient.**

We performed two simulations using both standard active inference and a modified active inference scheme where perception is performed using natural gradient descent. We compare the information length of belief updating of each scheme, with 128 agents across 24 trials and using a standard step size of $\varepsilon$=0.25. The first paradigm simulates a rat in a T-Maze (see *panel 3a*). The T-Maze has a reward which is placed either in the right or left upper arm (in red). The bottom arm contains a cue (in blue) that specifies the location of the reward. The rat's initial location is the middle of the T-Maze, as shown in the picture. The initial conditions are specified such that the rat believes that it will claim the reward and avoid the upper arm that does not have the reward. The optimal strategy consists of collecting the cue (exploration) and then using this information to collect the reward (exploitation). To achieve this, the rat must infer its location and the configuration of the maze, in addition to the route it will take. In this simulation, there are two hidden states, which correspond to the location of the rat in the T-Maze and the location of the reward, respectively. For details of this paradigm, the generative model and ensuing simulation see [57]. The second paradigm was a more complex simulation of abstract rule learning, with the same sort of generative model and belief updating mechanism as the first simulation. For more details on the paradigm see [39]. The purpose of including this is that it includes a hidden state dimension with three possible alternatives, facilitating a simple visual representation of the associated simplex in *panel 3d*. The histograms show the information length of belief updating in active inference (in red) and natural gradient (in blue) during the T-Maze and abstract rule learning tasks (resp. *panel 3b, 3e*). Specifically, this is the information length accrued by each agent at each trial, averaged across agents. One can see that the performance of both schemes as scored in terms of information length is almost identical across tasks. The reasons for systematic variation in information length across trials is that 1) the task configuration varied from trial to trial and 2) the generative models (i.e., representations of the environment) were themselves optimised (i.e., learned) over trials. The boxplots (resp. *panels 3c, 3f*) illustrate the difference in information length of the histograms (resp. *panels 3b, 3e*) by subtracting the information length of active inference from the information length of the natural gradient. In the first paradigm, active inference mostly takes shorter paths, which is why the boxplot's values are mostly positive. However, we obtain the opposite pattern in the second simulation. In addition, the differences in information length are marginal compared to the information length of each trial. Furthermore, both schemes perform equally well across both tasks on average. This suggests that the differences between the two schemes is small. *Panel 3d* shows an example of the belief trajectories taken during state estimation in abstract rule learning. The red trajectory is standard active inference, the blue is natural gradient descent, and the orange is the shortest path to the free energy minimum (i.e., the geodesic, see Appendix D). This example is not representative of the average and was chosen for purely illustrative purposes as the trajectories are very distinct, lengthy and do not coincide with the geodesic. The fact that both schemes take significantly longer paths than the geodesic was expected to occur in some trials as beliefs evolve to the free energy minimum myopically. Note that this apparent suboptimality was atypical in the tasks considered.

To summarise, our results suggest that state-estimation in active inference is a good approximation to natural gradient descent on free energy. Natural gradient descent follows the steepest descent of the objective in information space. Since the free energy optimised during state-estimation is convex (see Appendix A), this means that neural dynamics under active inference take short paths to the free energy minimum, which are the most computationally and metabolically efficient.

## Discussion

In the first part of this paper, we showed that neural dynamics under active inference are consistent with mean-field models neural population dynamics. This construct validity is further supported by the wide range of plausible electrophysiological responses that can be synthesised with active inference [57]. Yet, to fully endorse this view, the electrophysiological responses simulated during state-estimation need empirical validation. To do this, one would have to specify the generative model that a biological agent employs to represent a particular environment. This may be identified by comparing alternative hypothetical generative models with empirical choice behaviour and computing the relative evidence for each model (e.g., [119]). Once the appropriate generative model is found, one would need to compare the evidence for a few possible practical implementations of active inference, which come from various possible approximations to the free energy [27,120,121], each of which yields different belief updates and simulated electrophysiological responses. Note that of possible approximations to the free energy, the marginal approximation which was used in our simulations currently stands as the most biologically plausible [120]. Finally, one would be able to assess the explanatory power of active inference in relation to empirical measurements and compare it with other process theories.

In the second part of this paper, we showed that the neuronal process theory associated with active inference approximates natural gradient descent for state-estimation. Given that the natural gradient follows the direction of steepest descent of the free energy in information space [85] and that the free energy landscape at hand is convex (see Appendix A), this ensures that agent's beliefs reach the point of optimal inference via short trajectories in information space. This means that active inference entails neuronal dynamics that are both computationally and energetically efficient, an important feature of any reasonable process theory of the brain.

In the case of simulated (i.e., discretised) belief dynamics, active inference and natural gradient perform equally well on average. Performance is scored by the information length accrued during belief updating, which measures efficiency. In some cases, however, the belief trajectories taken by both schemes were significantly longer than the shortest path to the point of optimal inference. This is unsurprising since agents' beliefs move myopically to the free energy minimum. In short, our analysis suggests that biological agents can perform natural gradient descent in a biologically plausible manner. From an engineering perspective, this means that we can relate variational message passing and belief propagation, two inferential algorithms based on free energy minimisation [120,122,123], with the natural gradient.

A general point is that the tools furnished by information geometry are ideally suited to characterise and visualise inference in biological organisms as well as scoring its efficiency. This paves the way for further applications of information geometry to analyse information processing in biological systems.

## Conclusion

In the first part, we showed the consistency between the generic, first-principles account of brain function provided by active inference with the more detailed and empirically driven mean-field models of neural population dynamics.

Then, we demonstrated that perception under active inference approximates natural gradient descent on free energy. This suggests that the beliefs about states of the world of biological agents evolve approximately according to a steepest descent on free energy. Since the free energy landscape for discrete state-estimation is convex, the trajectories taken to the point of optimal inference are short, which incurs minimal computational and metabolic cost. This demonstrates the efficiency of neural dynamics under active inference, an important feature of the brain as we know it.

Further testing of active inference as a process theory of brain function should focus on extending the empirical evaluation of simulated neurophysiological responses.

## Software availability

The belief updating process described in this article are generic and can be implemented using standard routines (e.g., spm_MDP_VB_X.m). These routines are available as Matlab code in the SPM academic software: http://www.fil.ion.ucl.ac.uk/spm/. Examples of simulations can be found via a graphical user interface by typing DEM (e.g., DEM_demo_MDP_X.m for the T-Maze task [57], rule_learning.m for the artificial curiosity and abstract rule learning task [39]).

## Data accessibility

This article has no additional data.

## Authors' contributions

All authors made substantial contributions to conception, design and writing of the article; and approved publication of the final version.

## Competing interests

We have no competing interests.

## Funding

LD is supported by the Fonds National de la Recherche, Luxembourg (Project code: 13568875). TP is supported by the Rosetrees Trust (Award number: 173346). KF is funded by a Wellcome Trust Principal Research Fellowship (Ref: 088130/Z/09/Z).

## Appendix A. Convexity of the free energy

On discrete state-space generative models (e.g., partially observable Markov decision processes), the free energy optimised during state-estimation can be expressed as [7]:

$$\mathrm{F}(\mathbf{s}_{\pi 1},\ldots,\mathbf{s}_{\pi T}) = \sum_{\tau=1}^{T} \mathbf{s}_{\pi\tau} \cdot \ln \mathbf{s}_{\pi\tau} - \sum_{\tau=1}^{t} o_\tau \cdot \ln(A)\mathbf{s}_{\pi\tau} - \mathbf{s}_{\pi 1} \cdot \ln D - \sum_{\tau=2}^{T} \mathbf{s}_{\pi\tau} \cdot \ln(B_{\pi_{\tau-1}})\mathbf{s}_{\pi\tau-1}$$

In this context, the neuronal dynamics described in the paper are:

$$\dot{v}(\mathbf{s}_{\pi 1},\ldots,\mathbf{s}_{\pi T}) = -\nabla_{\mathbf{s}_{\pi\tau}} \mathrm{F}(\mathbf{s}_{\pi 1},\ldots,\mathbf{s}_{\pi T})$$
$$\mathbf{s}_{\pi\tau} = \sigma(v)$$

Here $\tau$ corresponds to time (which is discretised), $\mathbf{s}_{\pi\tau}$ corresponds to the beliefs about states at timepoint $\tau$, conditioned upon the fact that the agent is pursuing a certain sequence of actions $\pi$. The particular meaning of the other variables is not important for our purposes; the only important thing is that $B_{\pi_{\tau-1}}$ are matrices, whose components are strictly contained between zero and one, and logarithms are taken component-wise.

Recall that a sum of convex functions is convex. Furthermore,

- $x \mapsto x \ln x$ is convex in the interval $[0,1]$, which implies that $\sum_{\tau=1}^{T} \mathbf{s}_{\pi\tau} \cdot \ln \mathbf{s}_{\pi\tau}$ is convex.
- $-\sum_{\tau=1}^{t} o_\tau \cdot \ln(A)\mathbf{s}_{\pi\tau} - \mathbf{s}_{\pi 1} \cdot \ln D$ is a linear function, hence it is convex.
- $-\ln(B_{\pi_{\tau-1}})$ only has positive components, hence $-\sum_{\tau=2}^{T} \mathbf{s}_{\pi\tau} \cdot \ln(B_{\pi_{\tau-1}})\mathbf{s}_{\pi\tau-1}$ is a positive linear combination of polynomials of degree two, which is convex.

This implies that the free energy is convex.

## Appendix B. Fisher information metric tensor, information length and information distance

The Fisher information metric tensor is the canonical mathematical object that the enables computation of (a certain kind of) information distance on a statistical manifold. Technically, a metric tensor is a choice of symmetric, positive definite matrix at each point, that varies smoothly on the statistical manifold. This is equivalent to specifying an inner product at each point of the manifold and doing so smoothly.

Let $p(x\,|\,\mathbf{s})$ be a probability distribution parameterised by $\mathbf{s}$. The set of all possible choices of $\mathbf{s}$ is the statistical manifold associated with $p$, which we will denote by $M$. This is (in the case of classical probability distributions, which includes the scope of this paper) a smooth manifold, where each point corresponds to a certain parameterisation of the probability distribution, i.e., a (smooth) statistical manifold. We can then define the Fisher information metric tensor as

$$\mathbf{g}(\mathbf{s}) = \nabla_\theta^2 D_{KL}\left[p(x\,|\,\mathbf{s}) \,\|\, p(x\,|\,\theta)\right]\Big|_{\theta=\mathbf{s}}$$

This is an $n$-by-$n$ matrix where $n$ is the dimensionality of $\mathbf{s}$ and $\theta$. There exist other equivalent definitions [107,108].

This is nice, because a choice of an inner product at each point on the manifold enables to compute the length of tangent vectors. Let $v$ be such a tangent vector at a point $\mathbf{s}$, then its norm is given by

$$\|v\|_{\mathbf{g}} := \sqrt{v^T \mathbf{g}(\mathbf{s}) v}$$

This means that we can also compute the length of smooth curves. Let $\gamma:[0,1] \subset \mathbb{R} \to M$ be such a curve. Its information length is given by

$$\ell_{\mathbf{g}}(\gamma) := \int_0^1 \sqrt{\dot\gamma(t)^T \mathbf{g}\big(\gamma(t)\big)\dot\gamma(t)}\,\mathrm{d}t\,,$$

where $\dot\gamma := \frac{d\gamma}{dt}$.

We can trivially extend this definition to compute the information distance between points, say $\mathbf{s}$ and $\mathbf{s}'$. This is simply the information length of the shortest curve connecting the two points

$$d_{\mathbf{g}}(\mathbf{s},\mathbf{s}') = \inf_{\substack{\gamma:[0,1]\to M \\ \gamma(0)=\mathbf{s},\gamma(1)=\mathbf{s}'}} \ell(\gamma)\,.$$

Where, $\inf$ denotes the infimum of the quantity subject to the constraints in the subscript. Let us take a step back to see why these definitions are sensible.

Statistical manifolds are generally curved, therefore it is only possible to compute distances locally, by deforming the small region of consideration into a portion of Euclidean space. This is impractical and does not solve the problem of computing distances over larger scales. Even if one did so, one would recover a deformed version of the Euclidean distance, which would, generally speaking, not measure distance in terms of information. The raison d'être of the metric tensor is to allow the computation of distances on the manifold in a consistent way, and in our case consistently with the difference in Shannon information.

If one replaced $\mathbf{g}$ in the definitions above by the identity matrix (i.e., the metric tensor that is used implicitly in Euclidean space), one recovers the classical notion of length of a vector (i.e., the square root of the inner product), the classical notion of the length of a curve, namely

$$\ell(\gamma) := \int_0^1 \left\| \dot{\gamma}(t) \right\| \mathrm{d}t$$

The distance between two points is a little trickier as it involves proving that the shortest path between two points is the straight line when the metric tensor is the identity. This involves solving the geodesic equation (see Appendix D) for this metric tensor. Once this is done, inserting a straight line in the above equation returns the usual Euclidean distance.

## Appendix C. Fisher information metric tensor on the simplex

Suppose there are $n$+1 states $S = \{s_0, \ldots, s_n\}$. Then a categorical distribution $p(x \mid \mathbf{s})$ over those states is defined as $p(s_i \mid \mathbf{s}) := \mathbf{s}_i$. The statistical manifold of all possible parameters is the interior of the $n$-dimensional simplex which is defined as

$$\Delta^n := \{\mathbf{s} = (\mathbf{s}_0, \ldots, \mathbf{s}_n) \in \mathbb{R}^{n+1} \mid \mathbf{s}_i > 0, \sum_i \mathbf{s}_i = 1\}$$

The Fisher information metric tensor can be defined as $\mathbf{g}(\mathbf{s}) = \nabla_\theta^2 D_{KL}\left[ p(x \mid \mathbf{s}) \,\|\, p(x \mid \theta) \right]\Big|_{\theta=\mathbf{s}}$.

The KL-divergence between two categorical distributions is given by

$$\begin{aligned} D_{KL}\left[ p(x \mid \mathbf{s}) \,\|\, p(x \mid \theta) \right] &= \sum_{x \in S} p(x \mid \mathbf{s}) \log \frac{p(x \mid \mathbf{s})}{p(x \mid \theta)} \\ &= \sum_{i=0}^{n} p(s_i \mid \mathbf{s}) \log \frac{p(s_i \mid \mathbf{s})}{p(s_i \mid \theta)} \\ &= \sum_{i=0}^{n} \mathbf{s}_i \log \frac{\mathbf{s}_i}{\theta_i} \end{aligned}$$

We can take second derivatives

$$\frac{\partial}{\partial\theta_j}\frac{\partial}{\partial\theta_k}\left(\sum_{i=0}^{n}\mathbf{s}_i\log\frac{\mathbf{s}_i}{\theta_i}\right)=\delta_{jk}\frac{\mathbf{s}_k}{\theta_k^2}$$

Where $\delta_{jk}$ is the Kronecker delta. Finally,

$$\mathbf{g}(\mathbf{s})=\begin{matrix}\mathbf{s}_0^{-1} & 0 & \cdots & 0\\ 0 & \mathbf{s}_1^{-1} & \ddots & \vdots\\ \vdots & \ddots & \ddots & 0\\ 0 & \cdots & 0 & \mathbf{s}_n^{-1}\end{matrix}$$

Technical remark: since the statistical manifold of interest is $n$-dimensional, it is best to view this metric tensor as being defined on an $n$+1 dimensional neighbourhood of the simplex, e.g., the positive orthant of $\mathbb{R}^{n+1}$.

## Appendix D. Geodesics on the simplex

The aim of this section is to find the expression of the shortest path (in information length) between two points on the simplex.

As shown in Appendix C the metric tensor is given by

$$\mathbf{g}(\mathbf{s})=\begin{matrix}\mathbf{s}_0^{-1} & 0 & \cdots & 0\\ 0 & \mathbf{s}_1^{-1} & \ddots & \vdots\\ \vdots & \ddots & \ddots & 0\\ 0 & \cdots & 0 & \mathbf{s}_n^{-1}\end{matrix}$$

Let $\mathbf{s}^{(0)},\mathbf{s}^{(1)}$ be two points on the simplex. From standard differential geometry, the shortest path $\gamma$ between two points satisfies the geodesic equation:

$$\ddot{\gamma}_k+\sum_{i,j=0}^{n}\Gamma_{ij}^{k}(\gamma)\dot{\gamma}_i\dot{\gamma}_j\equiv 0$$

Where $\Gamma_{ij}^{k}$ are the Christoffel symbols of the Levi-Civita connection. These are real valued functions defined with respect to the metric:

$$\Gamma_{ij}^{k}:=\frac{1}{2}\sum_{r=0}^{n}\mathbf{g}^{kr}\left(\partial_j\mathbf{g}_{ri}+\partial_i\mathbf{g}_{rj}-\partial_r\mathbf{g}_{ij}\right)$$

In this expression $\mathbf{g}^{kr}$ is the $(k,r)$ entry of the inverse metric tensor $\mathbf{g}^{-1}$ and $\partial_j$ is a shorthand for $\frac{\partial}{\partial \mathbf{s}_j}$. In our case the only non-zero Christoffel symbols are given by

$$\Gamma^i_{ii}(\mathbf{s}) = -\frac{1}{2\mathbf{s}_i}$$

This means that each component of the geodesic must satisfy the equation

$$2\gamma_i\ddot{\gamma}_i - \dot{\gamma}_i^{\,2} \equiv 0$$

By inspection, one can see that the differential equation admits a polynomial solution of degree two. Solving with the boundary conditions $\gamma(0) = \mathbf{s}^{(0)}, \gamma(1) = \mathbf{s}^{(1)}$ and discarding those solutions that leave the positive orthant of $\mathbb{R}^{n+1}$ (c.f., last remark Appendix C) yields the expression of the geodesic:

$$\gamma(t) = \left((1-t)\sqrt{\mathbf{s}^{(0)}} + t\sqrt{\mathbf{s}^{(1)}}\right)^2$$

## Appendix E. Information distance on the simplex

The distance between two points on a statistical manifold is given by the information length of the shortest path (i.e., the geodesic) between the two. Given two points $\mathbf{s}^{(0)}, \mathbf{s}^{(1)}$ on the simplex, we have seen in Appendix D that the geodesic between these points is

$$\gamma(t) = \left((1-t)\sqrt{\mathbf{s}^{(0)}} + t\sqrt{\mathbf{s}^{(1)}}\right)^2$$

Furthermore, from Appendix B we have seen that the information distance between two points is the information length of the geodesic between them

$$d_{\mathrm{g}}(\mathbf{s}^{(0)}, \mathbf{s}^{(1)}) = \ell_{\mathrm{g}}(\gamma) = \int_0^1 \sqrt{\dot{\gamma}(t)^T \mathbf{g}\big(\gamma(t)\big)\dot{\gamma}(t)}\,\mathrm{d}t$$

Lastly, from Appendix C, the Fisher information metric tensor on the simplex is

$$\mathbf{g}(\mathbf{s}) = \begin{matrix} \mathbf{s}_0^{-1} & 0 & \cdots & 0 \\ 0 & \mathbf{s}_1^{-1} & \ddots & \vdots \\ \vdots & \ddots & \ddots & 0 \\ 0 & \cdots & 0 & \mathbf{s}_n^{-1} \end{matrix}$$

Therefore, expanding the expression inside the information distance

$$\dot{\gamma}(t)^T \mathbf{g}\big(\gamma(t)\big)\dot{\gamma}(t) = \sum_{i=0}^{n} \frac{\dot{\gamma}_i(t)^2}{\gamma_i(t)}$$

It is possible to show that $\frac{\dot{\gamma}_i(t)^2}{\gamma_i(t)}$ is constant for each $i$. One can do this by taking the derivative with respect to $t$ and noting that the result vanishes. This means that one can remove the integral and find a concise expression for the information distance:

$$\begin{aligned} d_{\mathbf{g}}(\mathbf{s}^{(0)},\mathbf{s}^{(1)}) &= \sqrt{\dot{\gamma}(0)^T \mathbf{g}\big(\gamma(0)\big)\dot{\gamma}(0)} \\ &= \sqrt{\sum_{i=0}^{n} \frac{\dot{\gamma}_i(0)^2}{\gamma_i(0)}} \\ &= \sqrt{4\sum_{i=0}^{n}\left(\sqrt{\mathbf{s}_i^{(1)}} - \sqrt{\mathbf{s}_i^{(0)}}\right)^2} \\ &= 2\left\|\sqrt{\mathbf{s}^{(1)}} - \sqrt{\mathbf{s}^{(0)}}\right\| \end{aligned}$$

This expression is compelling, since it relates the information distance on the simplex to the Euclidean distance on the *n*-dimensional sphere.